\documentclass[11pt]{article}
\usepackage{graphicx}
\usepackage{epsf}
\usepackage{epsfig}
\newcommand{\R}{{\sf R\hspace*{-0.94ex}%
\rule{0.15ex}{1.5ex}\hspace*{0.94ex}}}
\newcommand{\Z}{{\sf Z\hspace*{-0.94ex}%
\rule{0.15ex}{1.5ex}\hspace*{0.99ex}}}
\newcommand{\N}{{\sf N\hspace*{-0.99ex}%
\rule{0.15ex}{1.5ex}\hspace*{0.99ex}}}

\title{\bf Lattice fractional Laplacian and its continuum limit kernel on the finite cyclic chain}

\author{ { T.M.  Michelitsch$^{1}$\footnote{Corresponding author, e-mail~: michel@lmm.jussieu.fr}, B. Collet$^{1}$, A.F. Nowakowski $^{2,3}$, F.C.G.A Nicolleau$^{1,2,3}$ } \\ \\
$^1$ Sorbonne Universit\'{e}s\\
Universit\'{e} Pierre et Marie Curie, Paris 6\\
$^2$Institut Jean le Rond d'Alembert, CNRS UMR 7190 \\
4 Place Jussieu \\
75252 Paris cedex 05 \\
FRANCE\\ \\
$^{2}$
Department of Mechanical Engineering\\
Sir Frederick Mappin Building \\
Mappin Street \\
Sheffield \\
S1 3JD
United Kingdom \\ \\
$^{3}$ Sheffield Fluid Mechanics Group  \\
www.sheffield.ac.uk/fm \\
University of Sheffield\\
United Kingdom\\ \\ \\
{\it Accepted manuscript, to appear in Chaos, Solitons \& Fractals}}
\begin{document}
\maketitle
\paragraph{Abstract}
The aim of this paper is to deduce a discrete version of the fractional Laplacian in matrix form defined on the 1D periodic (cyclically closed) linear chain of finite length.
We obtain explicit expressions for this fractional Laplacian
matrix and deduce also its periodic continuum limit kernel. The continuum limit kernel gives an exact expression for the fractional Laplacian (Riesz fractional derivative) on the finite periodic string.
In this approach we introduce two material parameters, the particle mass $\mu$ and
a frequency $\Omega_{\alpha}$. The requirement of finiteness of the the total mass and total elastic energy in the continuum limit (lattice constant $h\rightarrow 0$) leads to scaling relations for the two parameters, namely
$\mu \sim h$ and $\Omega_{\alpha}^2\sim h^{-\alpha}$.
The present approach can be generalized to define lattice fractional calculus on periodic lattices in full analogy 
to the usual `continuous' fractional calculus.
\newline
PACS number(s): 05.50.+q, 02.10.Yn, 63.20.D-, 05.40.Fb
\newline
{\it Keywords~: Lattice fractional Laplacian, fractional Laplacian matrix, Lattice fractional calculus, fractional lattice dynamics, Riesz fractional derivative, discrete fractional Laplacian, cyclic chain, L\'evy lattice,
discrete fractional calculus, centered fractional differences, periodic Riesz fractional derivative, periodic fractional Laplacian, power-law matrix functions.}

\section{Introduction}

Many phenomena in nature are characterized by particle trajectories with irregular non-differentiable `complex' characteristics which often appear to be `similar' if one changes the scale. This property of self-similarity  
has as consequence that these phenomena cannot be decribed by
integer order partial differential equations. However, it has turned out that application of certain nonlocal 'fractional' operators on these trajectories may be well defined and hence these phenomena can be 
described by fractional partial differential equations, i.e. by differential equations of non-integer orders. This inevitable change of mathematical tools to describe such `anomalous' phenomena comes along with need of change of the goemetrical description of these 'fractal' 
trajectories. The traditional geometrical discription of integer order dimensions has to be given up and generalized to a description with non-integer order `fractal' dimensions. The need of a `fractionalized' description can occur 
with respect to space and time. 

The notion of fractal geometry introduced by Mandelbrot \cite{feder,mandelbr} taught us that the
classical idealizations for the trajectories of motions by smooth lines are rather rarely justified in natural structures. Instead nature chooses
irregular non-differentiable self-similar curves with a fractal non-integer dimension. It is today believed that fractional calculus in his numerous variants is the appropriate mathematical tool to
analyze such irregular motions.

Once a power law occurs in Fourier space, the Fourier transformed quantity is described naturally by a fractional operator.
There is recently a vast literature emerging to develop appropriate approaches to describe fractal phenomena in various physical contexts. 1D linear chains of infinite length leading to fractal dispersion relations were
analyzed in \cite{tara1,michel}. One direction of analysis suggests to model fractal domains embedded in the physical space by 
vector analysis developed for spaces of non-integer dimensions
\cite{tara2,tara3}, and see also the references therein. 
We emphasize that despite of the wide field of possible fractal applications of the approach to be developed, ``fractals'' are not the subject of the present paper.
The present paper is devoted to develop exact representations of fractional lattice Laplacian defined on finite periodic linear chains and to the analysis of its continuum limits. 
An overview on the relations between fractional calculus and fractal curves such as the Weierstrass Mandelbrot function has been presented by West \cite{west}.

Power law behavior which is naturally described by fractional calculus occurs in various
completely different contexts such as anomalous and turbulent diffusion, critical phenomena such as phase transitions, biological systems, the human economy,
and last but not least the present crisis of the financial system teaches us that complex systems such as the world economy does not obey gaussian statistics where extreme events are extremely seldom, however
they are governed by stable heavy tailed L\'evy distributions where extreme events are in the heavy tail still rare, but they are anyway much more likely as in gaussian cases.

There is a great variety of definitions for fractional integrals and derivatives depending on the function spaces in which they are defined, including Riemann, Liouville, Caputo, 
Gr\"unwald-Letnikow, Marchaud, Weyl, Riesz, Feller, among others, see e.g.
\cite{hilfer-2008,metzler,metzler2014,metzler-barkai,podlubny,samko,samko2003,riesz,tarafracrev,west}.

In the review article of Metzler and Klafter \cite{metzler} the random walk concept is applied in order to derive fractional equations 
of diffusion, diffusion-advection, and Fokker Planck type. In that article a broad overview of applications of fractional calculus on processes of anomalous diffusion is presented.
In the paper of Metzler et al. \cite{metzler-barkai} a fractional generalization of Fokker Planck equation is derived by analyzing the jump rates of a generalized Master equation. It can be said that
the fractional approach is inevitable to describe such anomalous phenomena. In another recent review article by the same authors
\cite{metzler2014} various 'anomalous' processes are discussed which are governed by fractional evolution equations, and a broad collection 
of systems is presented which exhibit anomalous processes such as L\'evy walks and -flights
with long-range correlations. In that article both, theoretical and experimental issues of anomalous transport processes governed by fractional dynamics are analyzed.

Whereas continuum fractional calculus is well developed, 
the lattice fractional calculus has rarely been considered. The development of lattice fractional calculus has become an important issue due to a variety of newly emerging applications, e.g. for the description of the dynamics on networks
\cite{riascos}. 
In two recent articles by Tarasov, fractional calculus on infinite lattices has been introduced \cite{tara4,tara5}. 
In \cite{tara5}  an analogue of vector fractional calculus is suggested. 
In that paper
long range interparticle interaction kernels are proposed defining fractional partial derivatives on lattices with power law Fourier transforms, 
similar to the kernels of continuum fractional derivatives. 
The difference of that approach and those to be introduced in the present paper is that we deduce from `fractional' elastic potentials the fractional Laplacian matrix on the {\it finite} cyclic chain.
To this end we start with an elastic potential generated 
by a quadratic form involving a power law matrix function 
(below relation (\ref{fracladiscc}))
of the discrete Born von Karman Laplacian matrix which is defined by (\ref{bvonkarma}). The fractional lattice approach suggested in \cite{tara4} holds on infinite bounded 
lattices. The approach introduced in the present paper includes both, infinite
and finite cyclic chains. 

The present paper aims to introduce a definition of fractional Laplacian which is analogue to the continuous version, being defined as an in general non-integer power 
of a ''Laplacian``, where the Laplacian is the
symmmetric centered second difference operator (Born-von Karman Laplacian matrix) well defined on the periodic finite chain. 
A fractional generalization of the Laplacians defined on cyclic networks have recently been analyzed by Riasos and Mateos \cite{riascos}, and the references therein. In that paper the construction of fractional Laplacian in general
networks is worked out and applied to cyclic networks. The study of Laplacian matrices generally is of importance in graph theory as it gives information on topological properties of 
the network \cite{riascos}. In this context the fractional Laplacian matrix to be introduced is of interest for the study of anomalous 
'fractional diffusion' processes taking place on the cyclic chain.

From an engineering point of view the analytical tools developed in the present paper can be of interest to improve the existing description in turbulent diffusion. 
Such fractional approaches could have impacts due to their capacity to describe irregular, erratic and `turbulent' phenomena, to improve aerodynamics models and performances of aerodynamic properties  in aerospace and car production engineering.
Since such erratic dynamic systems cannot be described by conventional
differential equations with spatial or temporal derivatives of integer orders having smooth and continuous solution characteristics.
The erratic dynamical characteristics of `turbulence trajectories' require to be described  by a another language, namely the
{\it fractional} language since fractional order derivatives of those trajectories are defined whereas their integer order derivatives do not exist \cite{zaehle}.

In the present paper we focus on the cases where the fractional operators occur in the space domain,
in the form of a fractionally generalized Laplace operator
$-(-{\tilde \Delta})^{\frac{\alpha}{2}}$ with positive power law index $\alpha>0$. In the continuum limit such systems are characterized by asymptotic 
$-k^{\alpha}$-power law behavior in Fourier space.
An example for a discrete model which yields in the continuum limit 
the 1D infinite space {\it fractional Laplacian} (Riesz fractional derivative)\footnote{For the continuum limit kernels to be deduced we use synonymously the terms
`fractional Laplacian' and `Riesz fractional derivative'.} has been developed in recent years \cite{michel,michel-fcaa,michel-ima2014}.

The present paper is organized as follows: First we define the discrete version of the $N\times N$-matrix of fractional Laplacian on the infinite linear chain as a limiting case of a cyclically closed (periodic) linear chain of 
identical $N\rightarrow \infty$ particles. 
$N$ indicates the number of identical particles of the periodic chain. 

We utilize the exact infinite chain limit expression and construct explicit expressions for the $N\times N$ fractional Laplacian matrix of the cyclically closed (periodic) {\it finite chain} where the number $N$ of
particles is finite and not necessarily large.
Further, by employing Stirling's asymptotic expression for $\Gamma$-functions for large arguments, we then obtain exact explicit expressions 
for the continuum limit kernels for the infinite 1D space and finite periodic string. 

The obtained fractional Laplacian matrices and kernels for the {\it infinite} chain and space, respectively, coincide with expressions given 
earlier in the literature. Compared to our recent analysis on the subject \cite{michel-per-frac}, the present paper aims to give a less technical but more didactical presentation of the subject.

\subsection{Fractional Laplacian matrix on the cyclic 1D chain}
\label{1Dcyclicchain}

It turns out that it is convenient to deduce first expressions for the fractional matrices on the infinite chain and then to use them to construct the fractional matrices on the finite cyclic chain.
First let us consider the case of a cyclically closed (periodic) linear chain of $N$ identical particles ($N$ not necessarily large)
and particle mass $\mu$. The particles are assumed to be equidistantly distributed along the chain with interparticle distance (lattice constant) $h$ and each mass point $p$ has equilibrium position at $0\leq x_p=ph\leq L-h$ ($p=0,..,N-1$) 
where $L$ denotes the length of the chain. Let us denote by $u_p=u(x_p)$ the displacement field of particle $p$.
Further we impose periodicity (cyclic closure of the chain) which is expressed by the equivalent notations
$u_p=u_{p+N}$ and $u(x_p)= u(x_p+L)$
for the displacement field for which we use the equivalent notations $u_p=u(x_p)$.
Reflecting the cyclicity of the chain we use cyclic index convention, namely $p \rightarrow p \,\,\, mod\, (N)  \in \{0,1,..,N-1\}$.
The cyclic chain can be imagined as a closed ring of $N$ identical particles
without ends.

For the fractional discrete approach to be developed it is convenient to introduce the shift operator operator $D(\tau) = e^{\tau\frac{d}{dx}}$ and its 
adjoint (inverse) operator $D^{\dagger}(\tau) = D(-\tau) = e^{-\tau\frac{d}{dx}}$ 
which act on sufficiently smooth fields $u(x)$ as $D(\pm\tau)u(x)= u(x\pm\tau)$, i.e. shifting arguments by $\pm\tau$, respectively.
For the sake of writing economy when we use
notation $D$ by skipping the argument, we mean the right hand sided next-neighbor particle shift operator $Du_p=:D(h)u(x_p)=u(x_p+h)= =u_{p+1}$, i.e. we utilize synonymously $D=:D(h)$
and its adjoint operator $D^{\dagger}=:D(-h)$ which indicates the left hand sided next-neighbor shift operator $D^{\dagger}u_p=u_{p-1}$.
The cyclicity of the chain is reflected by the periodicity of the shift operators
${\hat 1}=D(0)=D^{Nn}=D(nNh)=D(nL)$ ($n \in {\bf \Z}_0$). On the cyclic chain and in the limit of infinite length of the chain, the shift operator is unitary which is
reflected by $D^{-1}(nh)=D^{\dagger}(nh)=D(-nh)$. For the further analysis it is convenient to introduce the ``characteristic function" $f$
introduced as a scalar function with the properties \cite{michel-per-frac,michel-collet}
\begin{equation}
\label{charfu}
f(\lambda) >0 ,\hspace{1cm} 0<\lambda \leq 4 ,\hspace{1cm} f(\lambda=0)=0
\end{equation}
This condition $0<\lambda \leq 4$ is equivalent to elastic stability of the chain,
and $f(\lambda=0)=0$ describes the translational invariance of the chain (zero elastic energy for uniform translations).
We introduce the characteristic function with the dimensional units $[f] = {time}^{-2}$.
By utilizing a characteristic function with the properties (\ref{charfu}) the total elastic energy of stable 1D cyclic chain in the harmonic approximation can always be written in the form \cite{michel-collet}
\begin{equation}
\label{compfo}
V_f = \frac{\mu}{2}\sum_{p=0}^{N-1}u_p^*f(2{\hat 1}-D-D^{\dagger})u_p =: -\frac{1}{2} \sum_{p=0}^{N-1}\sum_{q=0}^{N-1} u_q^*
\Delta_f(|p-q|)u_p
\end{equation}
where
$\Delta_f(|p-q|)=-\mu f_{|p-q|}$ indicates the (negative semi-definite) Laplacian $N\times N$-matrix (discrete Laplacian operator) which fulfills also the periodicity condition $\Delta_f(|p-q|)=\Delta_f(|p-q+N|)$. 
In (\ref{compfo}) occurs the operator function $f(2-D-D^{\dagger})$ defined by the characteristic function $f$ having the properties (\ref{charfu}) with the generator matrix $2{\hat 1}-D-D^{\dagger}$ being the central symmetric Born von Karman Laplacian
having the $N\times N$ matrix representation
\begin{equation}
\label{matrixrep}
[2{\hat 1}-D-D^{\dagger}]_{pq}= 2\delta_{pq}-\delta_{p+1,q}-\delta_{p-1,q}
\end{equation}
where always cyclic index convention is always assumed. This includes the Kronecker symbol $\delta_{ij}={\hat 1}_{ij}$.
The characteristic function (\ref{charfu}) contains the entire
constitutive information of the chain-system. In the fractional lattice model to be developed we assume the characteristic function $f$ to be a power 
law function in the form of below relation (\ref{powerlaw}) and the goal is to determine explicit representations for the matrix elements of the fractional 
Laplacian matrix
(\ref{fracladiscc}) for infinite and finite cyclic chains.

It is straight-forward to see that for any characteristic function $f$ the eigenvectors on the periodic chain are the periodic Bloch vectors with the components
$\sim e^{i\kappa_lp}$ diagonalizing the second order difference operator
$(2-D-D^{\dagger})e^{i\kappa_lp}= 4\sin^2{\frac{\kappa_l}{2}} \, e^{i\kappa_lp} $ where $\kappa_l=\frac{2\pi}{N}l$ ($l=0,..,N-1$)
denotes $N$ non-dimensional Bloch wave numbers. It follows that they
are also diagonalizing on the cyclic chain any matrix function
$f(2-D-D^{\dagger})$ in (\ref{compfo}) which yields for its eigenvalues
the simple dispersion relation

\begin{equation}
\label{disprel}
\omega_f^2(\kappa_l) = f(\lambda = 4 \sin^2{\frac{\kappa_l}{2}}) \geq 0 ,\hspace{1cm} l=0,..,N-1
\end{equation}
where the zero eigenvalue occurs only for $l=0$ and $N-1$ positive eigenvalues are taken for $l=1,..,N-1$. These physically `good' properties are a consequence of (\ref{charfu}).
It follows from (\ref{charfu}) and (\ref{disprel})
that the characteristic matrix function $f(2-D-D^{\dagger})$ is a positive (semi-) definite $N\times N$ and self-adjoint (symmetric)
T\"oplitz matrix, i.e. of the form\footnote{Despite we utilize the notation $f_{pq}=f(|p-q|)$ for the matrix elements to indicate their dependence on $|p-q|$, this notation must never be confused with the 
$\lambda$-dependence $f(\lambda)$ indicated by (\ref{disprel}).} $f_{pq}= f(|p-q|)$.
The matrix elements fulfill periodicity and reflection-symmetry with respect to $\frac{N}{2}$,
namely\footnote{This symmetry is easily seen by putting $p=\frac{N}{2}+\chi$ and by accounting for
$f(|\frac{N}{2}+\chi|)=f(|-\chi-\frac{N}{2}+N|)=f(|\frac{N}{2}-\chi|)$.}
\begin{equation}
\label{percharfu}
\begin{array}{l}
f(|p|) = f(|p+nN|) ,\hspace{1cm} n \in {\bf \Z}_0 ,\hspace{1cm} 0\leq p \leq N-1\nonumber \\ \nonumber \\
f(|\frac{N}{2}+\chi| = f(|\frac{N}{2}-\chi|) = f(|nN+\frac{N}{2}-\chi|) ,\hspace{1cm}   \frac{N}{2}\pm \chi \in {\bf \Z}_0 , \hspace{0.5cm} n\in {\bf \Z}_0
\end{array}
\end{equation}
Due to periodicity, the points of reflection-symmetry repeat periodically being
located at $p_n=\frac{N}{2} +nN$ ($n \in {\bf Z}_0$).

\subsection{A lattice fractional approach for the 1D cyclic chain}

In this paragraph our goal is to develop rigerously the discrete fractional Laplacian on the infinite and finite cyclic chain being analogously defined as
the continuous version of the fractional Laplacian. The continuous fractional Laplacian in 1D formally is defined as a generally non-integer power of the Laplacian, namely as the 
negative (semi-definite) convolutional operator \cite{michel-ima2014}
\begin{equation}
 \label{fraccont}
 -\left(-\frac{d^2}{dx^2}\right)^{\frac{\alpha}{2}}\delta(x-x') ,\hspace{1cm} \alpha > 0
\end{equation}

The discrete Laplacian on the 1D cyclic chain can be defined consequently in analogous manner, namely
by replacing the 'continous' Laplace operator $\frac{d^2}{dx^2}$ by its discrete counterpart, i.e. by the Born von Karman Laplacian matrix
which is well defined on the cyclic chain by employing cyclic particle index convention \cite{michel-collet}

\begin{equation}
 \label{bvonkarma}
 {\tilde \Delta}_2 = (D+D^{\dagger}-2{\hat 1})
\end{equation}

We define hence the negative (semi-definite) `fractional Laplacian matrix' or synonymously `lattice fractional Laplacian' on the cyclic chain 
as fractional power law matrix function of the discrete Born von Karman Laplacian matrix (\ref{bvonkarma}) as
\begin{equation}
 \label{fracladiscc}
 \Delta_{\alpha} = -\mu\Omega_{\alpha}^2(-{\tilde \Delta}_2)^{\frac{\alpha}{2}} =  -\mu\Omega_{\alpha}^2\left(2{\hat 1}-D-D^{\dagger}\right)^{\frac{\alpha}{2}},\hspace{0.5cm} \alpha >0 
\end{equation}
where $\Omega_{\alpha}^2$ denotes a positive dimensional factor having physical dimension $sec^{-2}$. Whereas the continuous fractional Laplacian is for noninteger $\frac{\alpha}{2}$ a nonlocal convolutional kernel, its 
discrete counterpart defined on the cyclic chain of $N$ lattice sites appears as a non-diagonal $N\times N$ matrix of rank $N-1$. We notice that in the existing literature such as
\cite{tara5}, the lattice fractional calculus is 
considered by constructing the non-self adjoint fractional powers of derivatives $\sim (\frac{d}{dx})^{\alpha}$ on infinite lattices. In constrast
we deduce here the {\it fractional Laplacian matrix} defined by relation (\ref{fracladiscc}) as fractional powers of the simple Born von Karman Laplacian matrix (\ref{bvonkarma}) 
on the cyclic chain. The fractional Laplacian matrix (\ref{fracladiscc}) occurs as self-adjoint (symmetric) matrix of T\"oplitz structure. Further we demonstrate in section \ref{sec2} that the fractional Laplacian matrix of the cyclic chain
asymptotically yields in the continuum limit the Riesz fractional derivative kernel defined on the periodic string.

In order to demonstrate the physical meaning of the discrete fractional Laplacian (\ref{fracladiscc}), it is useful to note that this corresponds in the framework of the matrix 
function approach proposed in \cite{michel-collet}, to a power law characteristic function \cite{michel-per-frac}
\begin{equation}
\label{powerlaw}
f^{(\alpha)}(\lambda) = \Omega_{\alpha}^2 \,\lambda^{\frac{\alpha}{2}} , \hspace{1cm} \alpha >0
\end{equation}
where $\frac{\alpha}{2}$ denotes a positive, real (non-integer or integer) scaling index. 
In this paper we are especially interested in the {\it fractional} cases where $\frac{\alpha}{2} \notin {\bf N}_0$ is non-integer.
Note that positiveness of $\alpha$ is a consequence of the requirement that uniform translations of the chain
do not contribute to the elastic energy, and a consequence of the requirement $f^{(\alpha)}(\lambda=0)=0$ in (\ref{charfu}). As a consequence, the trivial value
$\alpha=0$ is physically forbidden \cite{michel-collet}.
The positiveness of $\alpha$ guarantees that the problem has physically ``good'' properties.
Then with (\ref{compfo}) and the power law characteristic function (\ref{powerlaw}) generates the `fractional elastic potential' in the form

\begin{equation}
\label{Valpha}
V_{\alpha} = \frac{\mu\Omega_{\alpha}^2}{2} \sum_{p=0}^{N-1}u_p^*(2-D-D^{\dagger})^{\frac{\alpha}{2}}u_p =:
\frac{\mu}{2} \sum_{p=0}^{N-1}\sum_{q=0}^{N-1}u_q^*f^{(\alpha)}(|p-q|)u_p
\end{equation}
which is a positive quadratic form in the displacements. 
In the next steps we evaluate the matrix elements in explicit form of the $N\times N$ power law matrix function

\begin{equation}
 \label{oppowfu}
 f^{\alpha}(2-D-D^{\dagger}) = \Omega_{\alpha}^2 (2-D-D^{\dagger})^{\frac{\alpha}{2}} = \Omega_{\alpha}^2 (-1)^{\frac{\alpha}{2}}\left(D(\frac{h}{2})-D(-\frac{h}{2})\right)^{\alpha} ,\hspace{2cm} \alpha >0
\end{equation}

Note that the elements of the fractional Laplacian matrix on the chain are connected with those of the characteristic matrix $f^{(\alpha)}(|p-q|)$ by
\begin{equation}
 \label{fraclaplamat}
 \Delta_{\alpha}(|p-q|)  = -\frac{\partial^2 V_{\alpha}}{\partial u_p\partial u_q} = -\mu f^{(\alpha)}(|p-q|)
\end{equation}
The $N$ eigenvalues (dispersion relation) of the power law operator (\ref{oppowfu}) is given by

\begin{equation}
\label{disprelat}
\omega_{\alpha}^2(\kappa_l) = f^{(\alpha)}\left(\lambda=4\sin^2{(\frac{\kappa_l}{2})}\right) = \Omega_{\alpha}^2\,2^{\alpha}\,|\sin{(\frac{\kappa_l}{2}})|^{\alpha} ,\hspace{0.5cm} \kappa_l=\frac{2\pi}{N}l , \hspace{0.15cm}  0 \leq l \leq  N-1
\end{equation}
with the only zero value for $l=0$ reflecting translational invariance of (\ref{Valpha}), and
$N-1$ positive values for $1\leq l \leq N-1$. The case $\alpha=2$ corresponds to the
classical Born von Karman chain having next neighbor particle springs, where (\ref{Valpha}) recovers the elastic energy of a Born-von-Karman next neighbor chain
and (\ref{disprelat})
then takes its well known classical dispersion relation $\omega_{2}^2(\kappa_l) =4\Omega_2^2\sin^2{(\frac{\kappa_l}{2}})$.

For the further analysis it is useful to write the $N\times N$ {\it fractional characteristic matrix function} $f^{(\alpha)}$ defined in (\ref{oppowfu}) in its
spectral representation
\begin{equation}
\label{alphacarfou}
 f_N^{(\alpha)}(|p-q|) = \sum_{l=0}^{N-1} \frac{e^{i\kappa_l(p-q)}}{N}\omega_{\alpha}^2(\kappa_l) =
 \frac{\Omega_{\alpha}^2}{N}\sum_{l=0}^{N-1} e^{i\kappa_l(p-q)} \left(4\sin^2{(\frac{\kappa_l}{2}})\right)^{\frac{\alpha}{2}}
\end{equation}

We observe that (\ref{oppowfu}) depends symmetrically on $D+D^{\dagger}$. As a consequence the characteristic matrix $f^{(\alpha)}$ together with
the fractional Laplacian matrix (\ref{fraclaplamat}) are self-adjoint and of T\"oplitz structure.
We can therefore conclude that (\ref{alphacarfou}) can be represented in terms of a ``centered'' infinite sum
\begin{equation}
\label{charfualpha}
\begin{array}{l}
\displaystyle \Omega_{\alpha}^2(2-D(h)-D(-h))^{\frac{\alpha}{2}} =  \sum_{p=-\infty}^{\infty}f_{N\rightarrow \infty}^{(\alpha)}(|p|)\, D(hp)
= f_{\infty}^{(\alpha)}(0)+\sum_{p=1}^{\infty}f_{\infty}^{(\alpha)}(|p|)\left\{D(ph)+D(-ph)\right\} \nonumber \\ \nonumber \\
\displaystyle \Omega_{\alpha}^2(2-D-D^{\dagger})^{\frac{\alpha}{2}} u_n = f_{\infty}^{(\alpha)}(0)u_n+\sum_{p=1}^{\infty}f_{\infty}^{(\alpha)}(|p|)\left\{u_{n-p}+u_{n+p}\right\}
\end{array}
\end{equation}
where this expression holds for the infinite chain ($N\rightarrow \infty$). The goal is now to evaluate the matrix elements
$f^{(\alpha)}(|p|)$ in explicit form, firstly {\bf (i)} $f_{\infty}^{(\alpha)}$ for the infinite chain $N\rightarrow\infty$, and secondly {\bf (ii)} $f_N^{(\alpha)}$ for the
finite cyclic chain when $N$ is arbitrary but not necessarily large.

\subsection{(i) Discrete fractional Laplacian on the infinite chain $N\rightarrow\infty$}
\label{infinitchain}

The spectral representation (\ref{alphacarfou}) can asymptotically be written as an integral (where the T\"oplitz structure of this matrix allows to put $|p-q| \rightarrow |p|$)

\begin{equation}
\label{fractlattice}
f_{\infty}^{(\alpha)}(|p|) = \frac{\Omega_{\alpha}^2}{2\pi}\int_{-\pi}^{\pi}e^{i\kappa p}\left(4\sin^2{\frac{\kappa}{2}}\right)^{\frac{\alpha}{2}} {\rm d}\kappa =\Omega_{\alpha}^2 \frac{2^{\alpha + 1}}{\pi}\int_{0}^{\frac{\pi}{2}}
\sin^{\alpha}(\varphi)\cos{(2p\varphi}){\rm d}\varphi
\end{equation}
where we used for $N\rightarrow \infty$ the asymptotics ${\rm d\kappa} \sim \frac{2\pi}{N}$ and $-\pi\leq \kappa_l=\frac{2\pi}{N}l \rightarrow \kappa \leq \kappa$.

The $p^{th}$ element of the infinite chain fractional characteristic matrix (where $p=|p| \neq 0$) then can be written as (Appendix eqs. (\ref{fractlatticeb})-(\ref{matrixele}))
\begin{equation}
\label{matrixeleb}
f^{(\alpha)}(|p|) = \Omega_{\alpha}^2\frac{2^{\alpha}}{\sqrt{\pi}}\frac{\frac{(\alpha-1)}{2}!}{(\frac{\alpha}{2}+p)!}(-1)^p\prod_{s=0}^{p-1}(\frac{\alpha}{2}-s) ,\hspace{1cm} p\neq 0
\end{equation}
which is obtained as (Appendix eqs. (\ref{matrixele}) ff.)
\begin{equation}
\label{matrixelei}
f_{\infty}^{(\alpha)}(|p|) =
\Omega_{\alpha}^2\,\frac{\alpha!}{\frac{\alpha}{2}!(\frac{\alpha}{2}+|p|)!}(-1)^p\prod_{s=0}^{|p|-1}(\frac{\alpha}{2}-s)  =
f_{\infty}^{(\alpha)}(0) (-1)^p\, \prod_{s=0}^{|p|-1}\frac{(\frac{\alpha}{2}-s)}{(\frac{\alpha}{2}+s+1)}  ,\hspace{1cm} p\neq 0
\end{equation}
where the diagonal element ($p=0$) yields 
\begin{equation}
\label{diag}
f_{\infty}^{(\alpha)}(0) = \Omega_{\alpha}^2\,\frac{2^{\alpha}}{\pi}\int_0^1 \xi^{\frac{\alpha-1}{2}}(1-\xi)^{-\frac{1}{2}}{\rm d}\xi = \Omega_{\alpha}^2\, \frac{2^{\alpha}}{\pi} \frac{(\frac{\alpha-1}{2})!(-\frac{1}{2})!}{\frac{\alpha}{2}!} =
\Omega_{\alpha}^2\,\frac{\alpha !}{\frac{\alpha}{2}!\frac{\alpha}{2}!} >0
\end{equation}
which necessarily is positive since because of the T\"oplitz structure being related with the trace $f^{(\alpha)}(0)=\frac{1}{N}Tr(f^{\alpha})$ of the positive semi-definite matrix $f^{(\alpha)}$. 

In order to write these expressions more compactly we introduce
generalized factorial function defined by \cite{abramo}
\begin{equation}
\label{gammafu}
\beta !=: \Gamma(\beta+1) , \hspace{1cm}, \Re(\beta) \neq -1,-2,..,-n,.. (n\in{\bf \N})
\end{equation}
where $\Gamma(..)$ denotes the $\Gamma$-function defined for complex $\beta$ except negative integers. The factorial function $\beta!$ of (\ref{gammafu})
has singularities at negative integers and obeys the recursion relation
\begin{equation}
 \label{furela}
 \beta! = \beta(\beta-1)!
\end{equation}

If $\beta$ takes integer values, (\ref{gammafu}) recovers the usual definition of the factorial. 
For $\Re(\beta) >-1$ (\ref{gammafu}) has the integral representation $\beta! = \int_0^{\infty}t^{\beta}e^{-t}{\rm d}t$ 
and for $\Re(\beta)-n < -1 (\Re(\beta)\neq -1, -2,-3,..), n\in {\bf \N}$ it can be generated with the recursion (\ref{furela}) and the integral representation
\begin{equation}
 \label{conti}
 (\beta-n)! = \frac{1}{\displaystyle \prod_{s=0}^{n-1}(\beta-s)!}\,\,\int_0^{\infty}e^{-\tau}\tau^{\beta}{\rm d}\tau ,\hspace{0.5cm} -1 <\beta< 0
\end{equation}
being defined for any negative non-integer real part argument $\Re(\beta)-n \notin -1, -2,-3,..,-s,..(s\in {\bf \N}) $.

In order to give (\ref{matrixelei}) a more convenient representation, the product occuring in that expression can be expressed by ($p\neq 0$)
\begin{equation}
 \label{forthepro}
  \prod_{s=0}^{|p|-1}(\frac{\alpha}{2}-s) = \frac{\frac{\alpha}{2}!}{(\frac{\alpha}{2}-|p|)!} 
\end{equation}
So the matrix elements (\ref{matrixelei}), (\ref{diag}) can compactly be written as $\forall p \in {\bf \Z}_0$
\begin{equation}
\label{generalizationalf}
f_{\infty}^{(\alpha)}(|p|) = \Omega_{\alpha}^2 \,(-1)^p\, \frac{\alpha!}{(\frac{\alpha}{2}-p)!(\frac{\alpha}{2}+p)!}
= \Omega_{\alpha}^2 \,(-1)^p\, \left(\begin{array}{l} \hspace{0.4cm} \alpha \nonumber \\  (\frac{\alpha}{2}+p) \end{array}\right)
\end{equation}
where we introduced the generalized binomial coefficients as follows

\begin{equation}
\label{generalizedbinomi}
\left(\begin{array}{l} \xi \nonumber \\ \xi_1 \end{array}\right) = \left(\begin{array}{l} \xi \nonumber  \\ \xi_2 \end{array}\right) =
\frac{\xi!}{\xi_1 !\xi_2 !} ,
\hspace{0.3cm} \xi=\xi_1+\xi_2 
\end{equation}
where the $\xi_i$ can take any nonintegers and any non-negative integers ($\xi\in \R, \xi_i \notin -1,-2,..$).
Due to the symmetry of (\ref{generalizedbinomi}) with respect to $\xi_1 \leftrightarrow \xi_2$
we recognize that the right hand side of (\ref{generalizationalf}) is symmetric with respect to $p\leftrightarrow -p$, so indeed $f_{\infty}^{(\alpha)}(|p|)$ depends only
on the absolute value $|p|$ reflecting the T\"oplitz structure of the matrix (\ref{charfualpha})$_1$.

We further notice that (\ref{generalizationalf}) is well defined for any $p \in {\bf \Z}_0$ when the analytically continued definition of the $\Gamma$-function is utilized.
We can further rewrite (\ref{generalizationalf}) when employing Euler's reflection formula in the
form\footnote{A detailed detailed derivation can be found in \cite{michel-per-frac}.}

\begin{equation}
\label{matrixformii}
f_{\infty}^{(\alpha)}(|p|) = -\Omega_{\alpha}^2\frac{\Gamma(\alpha+1)}{\pi}\sin{(\frac{\alpha\pi}{2})}\frac{\Gamma(p-\frac{\alpha}{2})}{\Gamma(\frac{\alpha}{2}+p+1)}
\end{equation}
This representation is especially useful for the asymptotic analysis for $|p| >>1$ which is important to deduce the power law continuum limit kernel
which is performed subsequently.

The matrix elements with $f^{(2m)}(|p|> m)=0 $ for integer $\frac{\alpha}{2}=m \in {\bf \N}$ are vanishing which 
also is reflected by the vanishing of $\sin{(\frac{\alpha\pi}{2})}$ for integers $\frac{\alpha}{2} = m \in {\bf \N}$ in (\ref{matrixformii})).
This demonstrates the localization of the fractional matrix for integer orders $\frac{\alpha}{2} = m  \in {\bf \N}$ where the non-zero elements (\ref{generalizationalf}) 
taking the integer binomial form
\begin{equation}
 \label{binomi}
 f_{\infty}^{(\alpha=2m)}(|p|) = \Omega_{2m}^2 \,(-1)^p\frac{(2m)!}{(m+p)(m-p)!}
\end{equation}

For non-integer $\frac{\alpha}{2} \notin {\bf \N}$, all elements $f^{\alpha}(|p|)$ are non-vanishing,
reflecting the {\it non-locality} of the fractional characteristic matrix. Representation (\ref{matrixformii}) is in accordance with the expression reported earlier \cite{Zoia2007}.
The negative semi-definite lattice fractional Laplacian matrix is then obtained from (\ref{generalizationalf}), (\ref{matrixformii}) by (\ref{fraclaplamat}). 
A more detailed discussion of the properties of $f_{\infty}^{(\alpha)}(|p|)$ is given in our recent paper \cite{michel-per-frac}.

In this paragraph (i) (Eqs. (\ref{charfualpha})-(\ref{binomi})) we developed exact expressions for the discrete {\it fractional Laplacian} defined by (\ref{fraclaplamat}) together with (\ref{oppowfu})
for the {\it infinite chain} (limiting case $N\rightarrow\infty$).
Now in the subsequent paragraph (ii) we deduce exact expressions for the fractional Laplacian matrix where 
we account for the finiteness of the cyclic chain, i.e. the particle number $N$ is arbitrary and not necessarily large.

\subsection{ (ii) Construction of the discrete fractional Laplacian on the finite cyclically closed chain}

In this paragraph we deduce by using above infinite chain results exact expressions for the components of $N\times N$-fractional
Laplacian matrix for the {\it finite cyclic chain} when $N$ is arbitrary and not necessarily large.
To this end we utilize the Fourier inversion of (\ref{fractlattice}) equivalent to the fact that Bloch waves $u_p \sim e^{i\kappa p}$ are
eigenvectors where for the infinite chain the Bloch wave numbers are continuous within the first Brillouin zone $-\pi \leq \kappa \leq \pi$ and $2\pi$-periodically continued, thus we have
with (\ref{disprel}) the general relation which holds the infinite chain limit $N \rightarrow \infty$

\begin{equation}
 \label{infchainspec}
 \sum_{p=-\infty}^{\infty} f_{\infty}(|p|)e^{i\kappa p} = f_{\infty}(p=0) + 2 \sum_{p=1}^{\infty} f_{\infty}(|p|)\cos{(\kappa p)} = \omega_f^2(\kappa)
\end{equation}
for any (also non-fractional) characteristic matrix $f_{\infty}$. We notice that in the infinite chain limit the Bloch wave numbers $\kappa_s=\frac{2\pi}{N}s \rightarrow \kappa$ become a quasi-continuous variable
so that relation (\ref{infchainspec}) holds indentically in a entire interval of finite length $-\pi \leq \kappa \leq \pi$.
Choosing now $\kappa =\kappa_l=\frac{2\pi}{N}s$ ($s=0,..,N-1$) where $\kappa_l$ denotes the Bloch wave number of a {\it finite chain} where $N$ is finite, relation (\ref{infchainspec}) holds as well and writes
\begin{equation}
 \label{finitekappal}
 \omega_f^2(\kappa=\kappa_l) =  \sum_{p=-\infty}^{\infty} f_{\infty}(|p|)e^{i\kappa_l p} = \sum_{p=0}^{N-1}e^{i\kappa_l p}\sum_{s=-\infty}^{\infty}f_{\infty}(|p+sN|)
 ,\hspace{1cm} \kappa_l= \frac{2\pi}{N}l ,\,\, l=0,..,N-1
\end{equation}
where we rearranged the terms by accounting for the $N$-periodicity of the phase factors $e^{i\kappa_l (p+sN)} = e^{i\kappa_l p}$ ($s\in {\bf \Z}_0$). Denoting by $f_N$ the finite chain characteristic matrix,
we find that

\begin{equation}
 \label{invertrel}
 \begin{array}{l}
\displaystyle  \sum_{l=0}^{N-1} \frac{e^{i\kappa_l p}}{N} \omega_f^2(\kappa=\kappa_l)= f_{N}(|p|) \nonumber \\ \nonumber \\
\displaystyle f_N(|p|)= \sum_{s=-\infty}^{\infty}f_{\infty}(|p+sN|) = f_{\infty}(|p|)+
 \sum_{s=1}^{\infty}\left\{ f_{\infty}(|p+sN|)+ f_{\infty}(|p-sN|)\right\} ,\hspace{1cm} p=0,..,N-1
 \end{array}
\end{equation}

This general expression relates the infinite chain characteristic matrix $f_{\infty}$ with the finite cyclic chain characteristic matrix $f_N$.
We observe further the asymptotic relation
\begin{equation}
 \label{infifini}
 \lim_{N\rightarrow\infty} f_N(|p|) = f_{\infty}(|p|)
\end{equation}

So we can write with (\ref{generalizationalf}) for the fractional characteristic matrix function of the finite cyclic chain the relation

\begin{equation}
\label{finiteelem}
\begin{array}{l}
\displaystyle f^{(\alpha)}_N(|p|) = \sum_{n=-\infty}^{\infty}f^{(\alpha)}_{\infty}(|p-nN|) = \displaystyle f^{(\alpha)}_{\infty}(|p|)+\sum_{n=1}^{\infty}
\left\{f^{(\alpha)}_{\infty}(|p+nN|)+ f^{(\alpha)}_{\infty}(|p-nN|)\right\}  \nonumber \\ \nonumber \\
\displaystyle \hspace{1.5cm} = \Omega_{\alpha}^2 (-1)^p \, \left(\begin{array}{l}
 \hspace{0.2cm} \alpha \nonumber \\
\frac{\alpha}{2} +p
\end{array}\right) + \Omega_{\alpha}^2\sum_{n=1}^{\infty}(-1)^{p+nN}\left\{\left(\begin{array}{l}
 \hspace{0.2cm} \alpha \nonumber \\
\frac{\alpha}{2} +p +nN
\end{array}\right)+\left(\begin{array}{l}
 \hspace{0.2cm} \alpha \nonumber \\
\frac{\alpha}{2} +p -nN
\end{array}\right) \right\} \hspace{0.4cm} p=0,..,N-1
\end{array}
\end{equation}

Relation (\ref{finiteelem}) is an exact relation which determines the finite chain fractional characteristic matrix $f^{(\alpha)}_N(|p|)$ by the fractional infinite chain characteristic
matrix of (\ref{generalizationalf}). The absolute convergence of series (\ref{finiteelem}) is guaranteed by the power law decay which is considered also in the subsequent paragraph.
The exact expression for the (negative semi-definite) fractional Laplacian matrix of the finite cyclic chain is
given by (\ref{fraclaplamat}) with (\ref{finiteelem}).

\section{Continuum limits of the discrete fractional model: 1D infinite space and finite
periodic string fractional Laplacian kernels}
\label{sec2}

Let us analyze the continuum limit kernel which is defined by the limiting case $h\rightarrow 0$ of the fractional
matrix elements first (\ref{matrixformii}) of the infinite cyclic chain.
We require for the continuum limit $h\rightarrow 0$ the following properties: \\
(i) The total mass of the chain $N\mu = M$ remains finite when the length of the chain $L=Nh$ is kept finite, i.e. does not depend on $h$ when $h\rightarrow 0$.\\
(ii) The total elastic energy of the chain remains finite when the length of the chain $L=Nh$ is kept finite,i.e. does not depend on $h$ when $h\rightarrow 0$.\\
This can be achieved when the two material constants $\mu$ and $\Omega_{\alpha}^2$ fulfill asymptotically for $h\rightarrow 0$
the following scaling relations \cite{michel-per-frac,michel-collet}
\begin{equation}
 \label{scalingrels}
 \Omega_{\alpha}^2(h) \sim A_{\alpha} h^{-\alpha} ,\hspace{2cm} \mu(h) \sim  \rho_0 h
\end{equation}
where $A_{\alpha}$ and $\rho_0$ are new positive constants independent of $h$.
The scaling of the mass holds when a homogeneous mass distribution with spatially constant mass density $\rho_0$ is assumed.
By assuming the scaling relations (\ref{scalingrels}) we see that the elastic energy (\ref{Valpha}) fulfills (ii), i.e. remains finite
in the continuum limit, namely

\begin{equation}
\label{Valphacontilim}
\lim_{h\rightarrow 0} V_{\alpha} = \lim_{h\rightarrow 0}\frac{\mu(h)}{2}\sum_{p=0}^{N-1}\sum_{q=0}^{N-1} u_pu_q^*f^{(\alpha)}(|p-q|)
=
\frac{\rho_0A_{\alpha}}{2} \int_0^L\int_0^Lu(x)u(x'){\cal K}^{(\alpha)}(|x-x'|){\rm dx}{\rm d x}'
\end{equation}
where ${\rm d}x \sim h$ and the continuum limit kernel ${\cal K}^{(\alpha)}(|x-x'|)$ is obtained when we take
into account Sterling's asymptotic
formula \cite{abramo} which holds asymptotically for
sufficiently large $\beta >>1$,
namely $\displaystyle \beta! \sim \sqrt{2\pi\beta} \,\,\frac{\beta^{\beta}}{{e^{\beta}}}$. Then
we obtain for $a,b << \beta$ finite a power law
\begin{equation}
\label{asympbet}
\frac{(\beta +a)!}{(\beta+b)!} \sim \beta^{a-b} \hspace{1cm} \beta >>1
\end{equation}

\subsection{(i) Infinite chain continuum limit: 1D infinite space Riesz fractional derivative}
Let us first consider the infinite space continuum limit where we assume $L \rightarrow \infty$. In the infinite space continuum limit the
elastic energy per unit length and the mass per unit length is assumed to remain finite.
This is achieved by assuming the same scaling relations (\ref{scalingrels}).

Introducing the spatial continuous variable $x=ph$ and $h\rightarrow 0$ (i.e. put in (\ref{matrixformii})
$0\leq |p|=\frac{|x|}{h} < \frac{L}{h} >>1$ by assming $|x|>x_{min}(h) \sim h^{\delta} const \rightarrow 0 $
and $|p|=\frac{x}{h} > p_{min}(h) =\frac{x_{min}}{h} \sim h^{\delta-1} >>1$ for $h$ small enough. This can be achieved for $0<\delta < 1$. By using then the expression (\ref{asympbet}),
we obtain for $|p|>>1$ for
(\ref{matrixformii})
the asymptotic representation
\begin{equation}
\label{asympmat}
\begin{array}{l}
\displaystyle f_{\infty}^{(\alpha)}(|p|)=\Omega_{\alpha}^2(-1)^p \left(\begin{array}{l}
\hspace{0.2cm} \alpha \nonumber \\
\frac{\alpha}{2} +p
\end{array}\right)= \nonumber \\ \nonumber \\
\displaystyle -\Omega_{\alpha}^2(h)\frac{\alpha!}{\pi}
\sin{(\frac{\alpha\pi}{2})}\frac{(p-\frac{\alpha}{2}-1)!}{(\frac{\alpha}{2}+p)!}
\sim -\Omega_{\alpha}^2(h)\,\,\frac{\alpha!}{\pi}\sin{(\frac{\alpha\pi}{2})} \,\, |p|^{-\alpha-1} =
-h A_{\alpha}\,\,\frac{\alpha!}{\pi}\sin{(\frac{\alpha\pi}{2})} \,\, |x|^{-\alpha-1}
\end{array}
\end{equation}
From (\ref{Valphacontilim}) follows\footnote{by taking into account that asymptotically for $h\rightarrow 0$ holds
$\displaystyle \sum_{p=0}^{N-1}\sum_{q=0}^{N-1}(..) = \frac{1}{h^2}\int_0^L\int_0^L(..){\rm d}x{\rm d}x'$} that the convolutional kernel
${\cal K}^{(\alpha)}_{\infty}$ is determined from

\begin{equation}
\label{asymp}
\rho_0A_{\alpha}{\cal K}^{(\alpha)}_{\infty}(|x|) = \lim_{h\rightarrow 0+} \frac{\mu(h)}{h^2} f_{\infty}^{(\alpha)}\left(\frac{|x|}{h}\right)
\end{equation}

\begin{equation}
\label{frackernel}
{\cal K}^{(\alpha)}_{\infty}(|x|) =
-\frac{\alpha!}{\pi}\sin{(\frac{\alpha\pi}{2})} \,\, |x|^{-\alpha-1} ,\hspace{0.5cm} x \neq 0
\end{equation}
Note that the fractional kernel (\ref{frackernel}) is the fractional kernel
of the infinite 1D space since it has been generated from the continuum limit of the infinite chain characteristic matrix (\ref{matrixformii}).

The (negative-semi-definite) {\it fractional Laplacian kernel} (Riesz fractional derivative) of the 1D infinite space simply is connected with
(\ref{frackernel}) by
\begin{equation}
\label{fractlaplkernel}
-\left(-\frac{d^2}{dx^2}\right)^{\frac{\alpha}{2}}\delta_{\infty}(x) = -{\cal K}_{\infty}^{(\alpha)}(|x|) =
\frac{\alpha !}{\pi}\frac{\sin{(\frac{\alpha\pi}{2})}}{ |x|^{\alpha+1}}  ,\hspace{0.5cm} x \neq 0
\end{equation}
The continuum limit kernel (\ref{fractlaplkernel}) has the dimensional units $cm^{-\alpha-1}$.
We emphasize that we are considering the infinite space and $\delta_{\infty}(x)$ indicates the infinite space Dirac's
$\delta$-function.
(\ref{fractlaplkernel}) is the well known expression for the kernel of infinite space fractional Laplacian
(e.g. \cite{riesz2,gorenflo,michel-ima2014} among many others)\footnote{Eq. (4.12) in \cite{riesz2}
uses a different sign convention. The expression there denotes the positive (semi-definite) operator $\left(-\frac{d^2}{dx^2}\right)^{\frac{\alpha}{2}}$ which is referred in that paper to as ``fractional Laplacian''.}. The hypersingular behavior at $x=0$ of
(\ref{fractlaplkernel}) can be removed by introducing a regularization in the distributional sense \cite{michel-ima2014}
\begin{equation}
\label{distri}
-\left(-\frac{d^2}{dx^2}\right)^{\frac{\alpha}{2}}\delta_{\infty}(x) = - \lim_{\epsilon\rightarrow 0+}\frac{\alpha !}{\pi}{\Re
\large\{\frac{i^{\alpha+1}}{(x+i\epsilon)^{\alpha+1}}}\large\}
\end{equation}

Further for integer $m=\frac{\alpha}{2}$, relation (\ref{distri}) takes indeed the distributional representation of integer order-Laplacian
\begin{equation}
\label{integerorder}
-\left(-\frac{d^2}{dx^2}\right)^{\frac{\alpha}{2}=m}\delta_{\infty}(x) = (-1)^{m+1}\frac{d^{2m}}{dx^{2m}} \lim_{\epsilon\rightarrow 0+}\frac{1}{\pi}\Re\{ \frac{i}{(x+i\epsilon)} \} = -(-1)^m\frac{d^{2m}}{dx^{2m}}\lim_{\epsilon\rightarrow 0+}\frac{1}{\pi}\frac{\epsilon}{(x^2+\epsilon^2)} ,\hspace{1cm} m\in {\bf \N_0}
\end{equation}
where $\displaystyle \delta_{\infty}(x)=\lim_{\epsilon\rightarrow 0+}\frac{1}{\pi}\frac{\epsilon}{(x^2+\epsilon^2)}$
is Dirac's infinite space $\delta$-function. Moreover, this relation corresponds to the distributional representation corresponding 
to the continuum limit $h^{-2m}(2-D-D^{\dagger})^m \sim \left(-\frac{d^2}{dx^2}\right)^m $, $m \in \N_0$.
This includes for $m=\frac{\alpha}{2}=1$ the continuum
limit of the classical infinite Born von Karman chain leading to local standard elasticity.

It is now only a small step to obtain the periodic string continuous limit when the finiteness of the chain (string)
$L=Nh$ is maintained.

\subsection{(ii) Periodic string continuum limit}

To obtain the fractional continuum limit kernel we consider directly the asymptotic expression (\ref{asympmat}) and apply it to the finite chain
fractional matrix $f_N^{(\alpha)}(|p|)$ of (\ref{finiteelem}) by
utilizing the asymptotic formula (\ref{asympbet}). We then arrive in the same asymptotic way
at the negative semi-definite Riesz fractional derivative (Fractional Laplacian) continuum limit kernel defined on the $L$-periodic string\footnote{where $\delta_L(..)$ denotes the $L$-periodically continued
Dirac $\delta$-function.}

\begin{equation}
\label{contilimlapfinal}
\begin{array}{l}
\displaystyle  -\left(-\frac{d^2}{dx^2}\right)^{\frac{\alpha}{2}}\delta_L(x) = -{\cal K}_L^{(\alpha)}(|x|) = \frac{\alpha ! \sin{(\frac{\alpha\pi}{2})}}{\pi} \sum_{n=-\infty}^{\infty}\frac{1}{|x-nL|^{\alpha+1}}  \nonumber \\ \nonumber \\
\displaystyle \hspace{1.5cm} = \frac{\alpha !\sin{(\frac{\alpha\pi}{2})}}{\pi L^{\alpha+1}}\left\{-\frac{1}{|\xi|^{\alpha+1}}+
{\tilde \zeta}(\alpha +1,\xi) + {\tilde \zeta}(\alpha +1,-\xi) \right\} ,\hspace{1cm} \xi=\frac{x}{L} \nonumber \\ \nonumber \\
\displaystyle \hspace{1.5cm} = -\frac{\alpha !}{\pi} \lim_{\epsilon\rightarrow 0+} \Re\left\{\sum_{n=-\infty}^{\infty}
\frac{i^{\alpha+1}}{(x-nL+i\epsilon)^{\alpha+1}}\right\} \nonumber \\ \nonumber \\ \displaystyle \hspace{1.5cm} =\frac{\alpha !}{\pi L^{\alpha+1}} \lim_{\epsilon\rightarrow 0+}
\Re \left\{\ i^{\alpha+1} \left(  \frac{1}{(\xi+i\epsilon)^{\alpha+1}}
-\zeta(\alpha+1,\xi+i\epsilon)-\zeta(\alpha+1,-\xi+i\epsilon) \right)\right\}
\end{array}
\end{equation}
For a more detailed derivation we again refer to \cite{michel-per-frac}.
In this expression for our convienience we introduced two kinds of Hurwitz $\zeta$-type functions, namely

\begin{equation}
\label{hurwitz}
{\tilde \zeta}(\beta,x)= \sum_{n=0}^{\infty}\frac{1}{|x+n|^{\beta}} \, ,\hspace{2cm} \zeta(\beta,x)=\sum_{n=0}^{\infty}
\frac{1}{(x+n)^{\beta}} ,\hspace{1.5cm} \Re\, \beta >1
\end{equation}

For $\alpha>0$ and $x\neq 0$ the series in (\ref{contilimlapfinal}) are absolutely convergent as good as the power
function integral  $\int_1^{\infty}\xi^{-\alpha-1}{\rm d}\xi$. The expressions (\ref{contilimlapfinal})$_{3,4}$ represent the regularized distributional representation of the kernel where
(\ref{distri}) has been taken into account. In (\ref{contilimlapfinal})$_{3,4}$ we take into account that regularized representation
$\Re \frac{i^{\alpha+1}}{(\xi+i\epsilon)^{\alpha+1}} = \Re \frac{i^{\alpha+1}}{(-\xi+i\epsilon)^{\alpha+1}} $
is for $\epsilon\rightarrow 0+$ gives an even distribution with respect to $\xi$.
The hyper-singular representations (\ref{contilimlapfinal})$_{1,2}$ shows that ${\cal K}_L^{(\alpha)}$
has periodically repeating singularities at $x=nL$ ($n\in {\bf \Z}_0$). It is easy to see that $\displaystyle \lim_{L\rightarrow \infty}{\cal K}_L^{(\alpha)}(|x|) = {\cal K}_{\infty}^{(\alpha)}(|x|)$
recovering the infinite space kernel defined in (\ref{fractlaplkernel}) corresponding to relation (\ref{infifini}) on the discrete chain.
In the limiting case when $\frac{\alpha}{2}=m\in {\bf \N}$ takes integers, the kernel (\ref{contilimlapfinal}) takes the distributional representation defined on the $L$-periodic string of integer
orders of the conventional Laplacian,
namely
\begin{equation}
\label{integerperiodic}
\begin{array}{l}
\displaystyle -{\cal K}_L^{(\alpha=2m)}(|x|)= (-1)^{m+1}\frac{d^{2m}}{dx^{2m}} \sum_{n=-\infty}^{\infty} \lim_{\epsilon\rightarrow 0+}\frac{1}{\pi}\frac{\epsilon}{((x-nL)^2+\epsilon^2)} ,\hspace{1cm} \frac{\alpha}{2} = m\in {\bf \N_0}\nonumber \\ \nonumber \\
\displaystyle \hspace{3cm} =
 (-1)^{m+1}\frac{d^{2m}}{dx^{2m}} \sum_{n=-\infty}^{\infty}\delta_{\infty}(x-nL) = -\left(-\frac{d^2}{dx^2}\right)^{m}\delta_L(x)
\end{array}
\end{equation}
where $\delta_L(x)$ denotes the $L$-periodic Dirac's $\delta$-function. 
For $m=\frac{\alpha}{2}=1$ (\ref{integerperiodic}) recovers the continuum
limit of the classical $L$-periodic Born von Karman chain leading to standard elasticity.

\section{Conclusions}
We deduced the lattice fractional Laplacian in explicit form for the infinite and finite periodic chain where in the latter case the particle number $N$ is arbitrary and not necessarily large. 
For integer exponents $\frac{\alpha}{2} \in \N$ 
the integer order binomial 
representation of the centered discrete second difference operator is recovered (relation (\ref{binomi})).

Further we analyzed continuum limits of the lattice fractional Laplacian: The infinite space continuum limit yields the distributional representations of the well known 1D fractional Laplacian kernel (Riesz fractional derivative). The infinite space representation is utilized to deduce
the $L$-periodic kernel of the fractional Laplacian on the $L$-periodic string. For all these cases, infinite space and periodic string, respectively, the fractional Laplacians take in the special 
case of integer exponents $\frac{\alpha}{2} \in {\bf \N}_0$ the distributional representations of the corresponding
integer-orders of the Laplacian in terms of localized even integer order derivatives of $\delta$-functions (relation (\ref{integerperiodic})).
The $L$-periodic string fractional Laplacian kernel represents the periodic string continuum limit of the discrete fractional Laplacian matrix of the finite $N$-periodic chain. The exact representation of 
the periodic string fractional Laplacian kernel (\ref{contilimlapfinal}) is expressed in compact form
by Hurwitz type $\zeta$-functions. Especially this representation appears to be useful for
computational purposes.

Moreover, the discrete fractional approach developed in this paper can be generalized to construct in analogous manner the fractional Laplacians on finite $n$-dimensional periodic lattices ($n$-tori where $n=1,2,3,..$). 
For instance for a cubic $nD$ periodic lattice the generalization of (\ref{fracladiscc}) is defined by

\begin{equation}
 \label{fraclaplanD}
 \Delta_{\alpha}^{(nD)} = -\mu \Omega_{\alpha}^2 \left(2n{\bf 1} -\sum_{i=1}^n(D_i+D_i^{\dagger})\right)^{\frac{\alpha}{2}} ,\hspace{1cm} \alpha >0
\end{equation}
where $n=1,2,3,..$ denotes here the dimension of the physical space.
The
spectral representation of fractional lattice Laplacian of the infinite cubic lattice in $n$ dimensions has then the spectral representation
\begin{equation}
\label{outl}
\begin{array}{l}
\displaystyle \Delta_{\alpha}^{(nD)}(|p_1|,|p_2|,..,|p_n|) = -\mu f^{(\alpha)}(|p_1|,|p_2|,..,|p_n|) ,\hspace{1cm} p_i \in {\bf Z}_0 \nonumber \\ \nonumber \\
\displaystyle f^{(\alpha)}(|p_1|,|p_2|,..,|p_n|) = 
\Omega_{\alpha}^2 \frac{2^{\alpha+n}}{\pi^n}\int_0^{\frac{\pi}{2}}{\rm d}\varphi_1\cos{(2p_1\varphi_1)} \times.. \nonumber \\ \nonumber \\
\displaystyle ..\times\int_0^{\frac{\pi}{2}}{\rm d}\varphi_n\cos{(2p_n\varphi_n)}
\left(\sum_{j=1}^n\sin^2{(\varphi_j)}\right)^{\frac{\alpha}{2}}
\end{array}
\end{equation}
which is 
the $nD$ generalization of the 1D infinite chain limit (\ref{fractlattice}) by accounting for (\ref{fraclaplamat}). Definition (\ref{outl}) for the fractional Laplacian matrix of the 
infinite cubic lattice recovers for $n=1$
the infinite chain limit expressions (\ref{fractlattice}) ff., and can also be developed with the fractional approach holding
on general networks, recently proposed by Riascos and Mateos \cite{riascos}.
At present an explicit evaluation of (\ref{outl}) for $n>1$ is not available. 
The availability of discrete fractional Laplacian matrices on infinite or finite $n$-dimensional tori, the latter being the $n$-dimensional counterparts of 1D cyclic chains, is highly desirable
as it would open to develop random walker strategies to study fractional (anomalous) transport processes on finite lattices.
In those anomalous diffusional problems on $nD$ lattices, diffusion equations occur where the continuous fractional Laplacian kernel $-(-\Delta)^{\frac{\alpha}{2}}$ is replaced by lattice
fractional Laplacian matrices of the form (\ref{outl}).

There seems to emerge a quite huge open field `{\it fractional lattice dynamics}' which could be explored by employing such a discrete fractional approach.

\section{Acknowledgements}
We thank G\'erard A. Maugin and Valery M. Levin for inspriring discussions. This work has been performed in the framework of the ERCOFTAC SIG/42 
cooperation project and the sabbatical stay of F.C.G.A Nicolleau at Institut Jean le Rond d'Alembert. 
We thank the two anonymous reviewers who considerably helped improving the manuscript.

\newpage

\section{Appendix}

In this appendix we evaluate in details the important integral (\ref{fractlattice}) for the matrix elements of the fractional Laplacian matrix

\begin{equation}
\label{fractlatticeb}
f^{(\alpha)}(|p|) = \frac{\Omega_{\alpha}^2}{2\pi}\int_{-\pi}^{\pi}e^{i\kappa p}\left(4\sin^2{\frac{\kappa}{2}}\right)^{\frac{\alpha}{2}} {\rm d}\kappa =\Omega_{\alpha}^2 \frac{2^{\alpha + 1}}{\pi}\int_{0}^{\frac{\pi}{2}}
\sin^{\alpha}(\varphi)\cos{(2p\varphi}){\rm d}\varphi \hspace{0.5cm} \alpha >0 ,\hspace{0.5cm} p\in {\bf \Z}_0
\end{equation}
Introduce $\xi=\sin^2(\varphi)$ (${\rm d}\varphi =2^{-1}\left[\xi(1-\xi)\right]^{-\frac{1}{2}}{\rm d}\xi$) with $0 \leq \xi \leq 1$ and $\cos{\varphi} = \sqrt{1-\xi} \geq 0,\sin{\varphi}=\sqrt{\xi} \geq 0$ where $0\leq \varphi \leq \frac{\pi}{2}$.
Further let us put in the following deduction $p=|p|$. Then (\ref{fractlatticeb}) writes as

\begin{equation}
\label{matelwrites}
f^{(\alpha)}(|p|) = \Omega_{\alpha}^2\frac{2^{\alpha}}{\pi}\Re \int_{0}^{1} \xi^{\frac{\alpha}{2}-\frac{1}{2}} (1-\xi)^{-\frac{1}{2}}\left(\sqrt{1-\xi}+i\sqrt{\xi} \right)^{2p} {\rm d}\xi
\end{equation}

Then we utilize
\begin{equation}
\displaystyle \cos{(2p\varphi}) = \Re\{(\sqrt{1-\xi}+i \sqrt{\xi})^{2p} \} =\sum_{s=0}^p \frac{(2p)!}{(2s)!(2p-2s)!}(-1)^s\xi^s(1-\xi)^{p-s}
\end{equation}
where $\Re(..)$ denotes the real part of $(..)$. Further we account for
\begin{equation}
\label{factorials}
(2n)! = 2^{2n}n!\frac{(n-\frac{1}{2})!}{(-\frac{1}{2})!} \hspace{1cm} n\in {\bf \N}_0
\end{equation}
so that (\ref{matelwrites}) can be written as
\begin{equation}
\label{easytosee}
f^{(\alpha)}(|p|) = \Omega_{\alpha}^2 \frac{2^{\alpha}}{\sqrt{\pi}} (p-\frac{1}{2})! \times
\int_{0}^{1}\xi^{\frac{\alpha}{2}}\sum_{s=0}^p\frac{p!}{s!(p-s)!}(-1)^s
\frac{\xi^{s-\frac{1}{2}}}{(s-\frac{1}{2})!}\frac{(1-\xi)^{p-s-\frac{1}{2}}}{(p-s-\frac{1}{2})!}{\rm d}\xi
\end{equation}
where by utilizing the Leibniz rule for $\frac{d^p}{d\xi^p}\{\xi^{p-\frac{1}{2}}(1-\xi)^{p-\frac{1}{2}}\}$ we have
\begin{equation}
\label{productderivative}
\begin{array}{l}
\displaystyle
 \Re\{\frac{(\sqrt{1-\xi}+i \sqrt{\xi})^{2p}}{\sqrt{\xi(1-\xi)}}\} = (-\frac{1}{2})!(p-\frac{1}{2})!\sum_{s=0}^p\frac{p!}{s!(p-s)!}(-1)^s
\frac{\xi^{s-\frac{1}{2}}}{(s-\frac{1}{2})!}\frac{(1-\xi)^{p-s-\frac{1}{2}}}{(p-s-\frac{1}{2})!} \nonumber \\ \nonumber \\
\displaystyle \hspace{3.75cm} = \frac{(-\frac{1}{2})!}{(p-\frac{1}{2})!}\frac{d^p}{d\xi^p}\left\{\xi(1-\xi) \right\}^{p-\frac{1}{2}}
\end{array}
\end{equation}
Thus we obtain
\begin{equation}
\label{fractalatb}
f^{(\alpha)}(|p|) = \Omega_{\alpha}^2 \frac{2^{\alpha}}{\sqrt{\pi}}\frac{1}{(p-\frac{1}{2})!} \int_0^1\xi^{\frac{\alpha}{2}}\frac{d^p}{d\xi^p}\left\{\xi(1-\xi)\right\}^{p-\frac{1}{2}}{\rm d}\xi
\end{equation}
where $(-\frac{1}{2})!=\Gamma(\frac{1}{2})=\sqrt{\pi}$. Note that for the (in our model) forbidden exponent $\alpha=0$ the matrix element (\ref{fractalatb}) yields $f^{(\alpha=0)}(|p|) = \Omega_0^2\,\delta_{p0}$ in accordance with ${\hat 1}\Omega_{0}^2$ of
relation (\ref{alphacarfou}) which also can directly be seen from (\ref{fractlatticeb}).
Now consider
\begin{equation}
\label{considerintehral}
\int_0^1\xi^{\frac{\alpha}{2}}\frac{d^p}{d\xi^p}\left\{\xi(1-\xi)\right\}^{p-\frac{1}{2}}{\rm d}\xi = \left\{\xi^{\frac{\alpha}{2}}\frac{d^{p-1}}{d\xi^{p-1}}\left\{\xi(1-\xi)\right\}^{p-\frac{1}{2}} \right\}_0^1 -\frac{\alpha}{2}\int_0^1\xi^{\frac{\alpha}{2}-1}\frac{d^{p-1}}{d\xi^{p-1}}\left\{\xi(1-\xi)\right\}^{p-\frac{1}{2}}{\rm d}\xi
\end{equation}
The lowest relevant orders of the boundary term $\{..\}$ is behaving as $\sim \xi^{\frac{\alpha+1}{2}}(1-\xi)^{\frac{1}{2}} $ + higher order terms and is hence vanishing at the boundary $\xi=0,1$. Performing $n \leq p$ times partial integration yields boundary terms with
lowest powers in $\xi$ and $1-\xi$ of the form

\begin{equation}
\label{boundaryterms}
\sim \xi^{\frac{\alpha}{2}-(n-1)}\frac{d^{p-n}}{d\xi^{p-n}}\left\{\xi(1-\xi)\right\}^{p-\frac{1}{2}}\sim
\xi^{\frac{\alpha+1}{2}}(1-\xi)^{n-\frac{1}{2}} ,\hspace{1cm} 1\leq n\leq p
\end{equation}
which are vanishing at the boundary $\xi=0,1$. Performing this procedure $p$ times yields then
\begin{equation}
\label{ntimes}
\begin{array}{l}
\displaystyle \int_0^1\xi^{\frac{\alpha}{2}}\frac{d^p}{d\xi^p}\left\{\xi(1-\xi)\right\}^{p-\frac{1}{2}}{\rm d}\xi = \int_0^1\xi^{\frac{\alpha}{2}-p}\left\{\xi(1-\xi)\right\}^{p-\frac{1}{2}}{\rm d}\xi \times\, (-1)^p \prod_{s=0}^{p-1}(\frac{\alpha}{2}-s)\nonumber \\ \nonumber\\
\displaystyle = \int_0^1\xi^{\frac{\alpha -1}{2}}\left\{(1-\xi)\right\}^{p-\frac{1}{2}}{\rm d}\xi \times\, (-1)^p \prod_{s=0}^{p-1}(\frac{\alpha}{2}-s)\nonumber \\ \nonumber \\
\displaystyle = \frac{\frac{\alpha-1}{2}!(p-\frac{1}{2})!}{(\frac{\alpha}{2}+p)!}\times\, (-1)^p \prod_{s=0}^{p-1}(\frac{\alpha}{2}-s)
\end{array}
\end{equation}
where in the last line we have used
\begin{equation}
\label{Bfunction}
\int_0^1\xi^{\beta_1}(1-\xi)^{\beta_2}{\rm d}\xi =\frac{\beta_1!\beta_2!}{(\beta_1+\beta_2+1)!},\hspace{1cm} \Re\, \beta_i >-1
\end{equation}
With (\ref{ntimes})$_3$ we get for the matrix element (\ref{fractalatb}) where always $p=|p|$
\begin{equation}
\label{matrixele}
f^{(\alpha)}(|p|) = \Omega_{\alpha}^2\frac{2^{\alpha}}{\sqrt{\pi}}\frac{\frac{(\alpha-1)}{2}!}{(\frac{\alpha}{2}+p)!}(-1)^p\prod_{s=0}^{p-1}(\frac{\alpha}{2}-s)
\end{equation}
To get this into a more convenient form we consider (\ref{Bfunction}) for $\beta_1=\beta_2=\frac{\alpha-1}{2}$
and by introducing $\xi=\frac{1}{2}(1+\sqrt{\eta})$ (${\rm d}\xi= 2^{-2}\eta^{-\frac{1}{2}}{\rm d}\eta$).
Then we have
\begin{equation}
\label{Bfualp}
\frac{(\frac{(\alpha-1)}{2})!(\frac{(\alpha-1)}{2})!}{\alpha !} =
2^{-\alpha}\int_0^1(1-\eta)^{\frac{(\alpha-1)}{2}}\eta^{-\frac{1}{2}}{\rm d}\eta = 2^{-\alpha} \frac{(\frac{(\alpha-1)}{2})!}{\frac{\alpha}{2}!}(-\frac{1}{2})!
\end{equation}
which is known as duplication formula \cite{abramo}. It follows with $(-\frac{1}{2})!=\sqrt{\pi}$ that
\begin{equation}
\label{2alphapi}
\frac{\alpha !}{\frac{\alpha}{2}!} = \frac{2^{\alpha}}{\sqrt{\pi}}(\frac{(\alpha-1)}{2})!
\end{equation}
Plugging (\ref{2alphapi}) into (\ref{matrixele}) yields a more illuminating representation, namely
\begin{equation}
\label{matrixeleibb}
f^{(\alpha)}(|p|) = \Omega_{\alpha}^2\frac{\alpha!}{\frac{\alpha}{2}!(\frac{\alpha}{2}+p)!}(-1)^p\prod_{s=0}^{p-1}(\frac{\alpha}{2}-s)
\end{equation}
where the analytically extended representation of the $\Gamma$-function is employed with $\beta !=\Gamma(\beta+1)$ , ($\beta \notin -1, -2,..$).
Now we observe that we can write
\begin{equation}
\label{observation}
\prod_{s=0}^{p-1}(\frac{\alpha}{2}-s) = \frac{\frac{\alpha}{2}!}{(\frac{\alpha}{2}-p)!}
\end{equation}
Thus we can write
\begin{equation}
\label{matrixelecasei}
f^{(\alpha)}(|p|) = \Omega_{\alpha}^2(-1)^p\frac{\alpha!}{(\frac{\alpha}{2}-p!)(\frac{\alpha}{2}+p)!}
\end{equation}
which obviously is a generalization of the binomial coefficients including integer and non-integer $\alpha$.

\end{document}